\documentclass[11pt]{article}

\usepackage[latin1]{inputenc}
\usepackage{graphicx}
\usepackage{amsmath}
\usepackage{color}
\usepackage{graphicx}
\usepackage{hyperref}
\usepackage{comment}
\usepackage[mathscr]{euscript}
\usepackage{algorithm}
\usepackage[noend]{algpseudocode}
\usepackage{natbib}
\usepackage{amssymb}
\usepackage{bm}
\usepackage{multirow}
\usepackage{enumitem}
\usepackage{lscape}
\usepackage{numprint}
\usepackage{relsize}

\newcommand{\xxi}{\mbox{\boldmath{$\xi$}}}
\newcommand{\ggamma}{\mbox{\boldmath{$\gamma$}}}
\newcommand{\ddelta}{\mbox{\boldmath{$\delta$}}}

\newcommand{\R}{\bf R}
\newcommand{\I}{\bf I}
\newcommand{\uu}{\bf u}
\newcommand{\vv}{\bf v}
\newcommand{\yy}{\bf y}

\begin{document}

\title{\textbf{A scalable approach for short-term disease forecasting in high spatial resolution areal data}}

\author{Orozco-Acosta, E.$^{1,2}$, Riebler, A.$^{3}$, Adin, A.$^{1,2}$ and Ugarte, M.D.$^{1,2}$\\
\small {\textit{$^1$ Department of Statistics, Computer Sciences and Mathematics, Public University of Navarre, Spain.}} \\
\small {\textit{$^2$ Institute for Advanced Materials and Mathematics, InaMat$^2$, Public University of Navarre, Spain.}}\\
\small {\textit{$^3$ Department of Mathematical Sciences, Norwegian University of Science and Technology, Norway.}}\\
\small {$*$Correspondence to Mar\'ia Dolores Ugarte, Departamento de Estad\'istica, Inform\'atica y Matem\'aticas } \\
\small {Universidad P\'ublica de Navarra, Campus de Arrosadia, 31006 Pamplona, Spain.} \\
\small {\textbf{E-mail}: lola@unavarra.es }}
\date{}

\makeatletter
\pdfbookmark[0]{\@title}{title}
\makeatother

\maketitle

\begin{abstract}
Short-term disease forecasting at specific discrete spatial resolutions has become a high-impact decision-support tool in health planning. However, when the number of areas is very large obtaining predictions can be computationally intensive or even unfeasible using standard spatio-temporal models. The purpose of this paper is to provide a method for short-term predictions in high-dimensional areal data based on a newly proposed ``divide-and-conquer" approach. We assess the predictive performance of this method and other classical spatio-temporal models in a validation study that uses cancer mortality data for the 7907 municipalities of continental Spain. The new proposal outperforms traditional models in terms of mean absolute error, root mean square error and interval score when forecasting cancer mortality one, two and three years ahead. Models are implemented in a fully Bayesian framework using the well-known integrated nested Laplace (INLA) estimation technique.
\end{abstract}


\bigskip

\section{Introduction}
\label{sec:Intro}

Bayesian hierarchical models have been developed extensively to model area-level incidence or mortality data and to estimate their underlying spatial, temporal and spatio-temporal patterns. Traditionally, generalized linear mixed models including spatially and temporally structured random effects have been proposed for smoothing disease risks or rates by borrowing information from neighbouring areas and time periods. In addition, extensions of these hierarchical models have been also considered for forecasting of rare and non-communicable diseases in areal data. For example, \cite{assuncao2001diffusion} provide an extension of the parametric model of \cite{bernardinelli1995bayesian} to predict human visceral Leishmaniasis incidence rates in 117 health zones of a Brazilian municipality. \cite{etxeberria2014} extend the non-parametric models proposed by \cite{knorr2000} that include conditional autoregressive (CAR) priors for space and random walk (RW) priors for time for short-term cancer mortality risk predictions in the 50 provinces of Spain. Similar projections of cancer mortality risks using spatio-temporal P-spline models were also considered by \cite{ugarte2012} and \cite{etxeberria2015predicting}. In \cite{corpas2021autoregressive}, an enhancement of a previous autoregressive (AR) spatio-temporal model proposed by \cite{martinez2008autoregressive} was considered for five-year ahead forecasting of different cancer site mortality data in the 540 municipalities of the Valencian autonomous region of Spain.  \cite{etxeberria2022} predict incidence rates for rare and lethal cancers by borrowing strength from mortality data using spatio-temporal models with shared spatial and age components. Different extensions of APC (age-period-cohort) models including spatial random effects have been also proposed for the prediction of cancer mortality and incidence data (see \citealp{lagazio2003,schmid2004bayesian,papoila2014} or \citealp{etxeberria2017} among others). Very recently, several extensions of different CAR, AR and RW models have been also proposed for representing the geographical variation of risk processes that underlay the dynamic outbreaks of COVID-19 infection and related outcomes (see, for example, \cite{macnab2023adaptive} and the references therein).

All these models  perform well when the spatial domain has a limited number of areas. If the number of areas is very large, model fitting becomes computationally expensive or even unfeasible \citep{vanNiekert2021}, mainly due to the huge dimension of the spatio-temporal covariance/structure matrices and the high number of identifiability constraints \citep{schrodle2011,goicoa2018spatio}. However, forecasting short-term disease risks or rates in high-resolution areal data is very important to take high-impact decisions in health planning and addressing health inequalities \citep{utazi2019spatial,sartorius2021modelling}. Cancer mortality projections also play an important role in epidemiology, as they support the decision-making process for population intervention plans and health resource planning. According to the Spanish Statistical Institute, cancer was the second leading cause of mortality in Spain in 2021 after cardiovascular diseases (24.3\% and 22.8\%, respectively), being the first leading cause of death among the male population. The estimated direct cost of cancer in Spain for the year 2019 was more than 7,000 millions of euros, which represents about 10\% of Spanish health costs \citep{diaz2019}.

The aim of the current paper is to evaluate if the scalable Bayesian spatio-temporal modelling approach proposed by \cite{orozco2022} for estimating risks is also appropriate for short-term forecasting in high spatial resolution areal data. This methodology is based on a ``divide-and-conquer'' approach and has been shown to provide reliable risk estimates with a substantial reduction in computational time in comparison with classical spatio-temporal CAR models. Specifically, \cite{orozco2022} propose to divide the spatial domain into smaller subregions where independent models can be fitted simultaneously by using parallel or distributed computation strategies. To reduce the border effect in the risk estimates caused by the spatial partitions, neighbouring areas are added to each subdomain when fitting the models. Finally, the risk estimates from different sub-models are properly combined to obtain unique posterior marginal estimates of the risks for each areal-time unit. This approach has been also extended to high-dimensional multivariate spatial models to jointly analyse several disease outcomes \citep{vicente2022high}. However, it has not been checked yet in a forecasting framework.

The rest of the paper is structured as follows. 
Sections~\ref{sec:Models} and \ref{sec:Scalable_models} outline the methodology and shortly describe the different spatio-temporal models considered for short-term forecasting of cancer mortality. A validation study is presented in Section~\ref{sec:ValidationStudy} to assess and compare the predictive performance of the models. In Section~\ref{sec:DataAnalysis}, the proposed methodology is applied to project male lung cancer and overall cancer (all sites) mortality data by considering three-year ahead predictions in the 7907 municipalities of continental Spain. 
The paper concludes with a discussion.

\section{Bayesian spatio-temporal models}
\label{sec:Models}
Let $y_{it}$ and $n_{it}$ denote the observed number of cancer deaths and the corresponding number of population at risk in region $i=1, \ldots, N$ and time period (year) $t=1, \ldots, T$, respectively. Here, we assume that all regions are connected and that years are consecutive. We further assume that the observations are conditionally independent and model them as
\begin{eqnarray*}
\begin{array}{rcl}
y_{it}|\lambda_{it} & \sim & \text{Poisson}(\mu_{it}=n_{it} \cdot \lambda_{it}) \ \text{for} \ i=1,\ldots,N; \ t=1,\ldots,T,\\[1.ex]
\log{\mu_{it}} & \sim & \log{n_{it}} + \log{\lambda_{it}}
\end{array}
\end{eqnarray*}
where $\log{n_{it}}$ is an offset and $\lambda_{it}$ is the mortality rate in region $i$ at time $t$. Depending on the specification of $\log{\lambda_{it}}$ we define different models
that are all placed in a hierarchical Bayesian inference scheme. Inference is carried out using the well-known integrated nested Laplace approximation technique (INLA) \citep{rue2009approximate} implemented in the R-INLA package (\url{www.r-inla.org}).
INLA has become a very popular and user-friendly inference tool for spatial and spatio-temporal disease mapping models (see, e.g., \citealp{schrodle2011using,ugarte2014,macnab-2022}), and in many cases is both faster and more accurate than Markov chain Monte Carlo alternatives \citep{rue2017bayesian}.

\subsection{Classical Bayesian disease-mapping models}

The great variability inherent to classical risk estimation measures such as crude rates when analysing very small domains or low-populated areas, requires the use of statistical models to smooth risks borrowing information from spatial and temporal neighbours \citep{wakefield2007disease}.
%
%
Here, we assume a linear predictor of the form
\begin{equation}
\label{eq:M1}
\log{\lambda_{it}} = \beta_0 + \xi_{i} + \gamma_{t} + \delta_{it},
\end{equation}
where $\beta_0$ is an intercept representing the overall log-rate, $\xi_i$ is a spatial random effect that follows the so-called BYM2 prior distribution \citep{riebler2016intuitive}, $\gamma_t$ is a temporally structured random effect that follows a first-order random walk (RW1), and $\delta_{it}$ is a spatio-temporal random effect allowing for space-time interactions \citep{knorr2000}. All the components of this model can be formulated as Gaussian Markov random fields \citep{rue2005gaussian} and prior densities can be written according to some structure matrices.

The BYM2 model for the spatial random effect ${\xxi}=(\xi_1,\ldots,\xi_N)^{'}$ is expressed as
\begin{equation*}
{\xxi} = \frac{1}{\sqrt{\tau_{\xi}}} \left(\sqrt{\phi}{\uu}_{\ast} + \sqrt{1-\phi}{\vv} \right),
\end{equation*}
where $\tau_{\xi}$ is a precision parameter, ${\uu}_{\ast} \sim \mathcal{N}(\bm{0}, {\R}_{\ast}^{-})$ is the scaled intrinsic CAR model with ${\R}_{\ast}^{-}$ representing the generalised inverse of the standardized neighbourhood structure matrix ${\R}_{\ast}$ (see \citealp{sorbye2014scaling}), ${\vv}\sim \mathcal{N}(\bm{0}, {\I}_N)$ is a vector of unstructured random effects, and $\phi \in [0,1]$ is a parameter that weights the regional variability between the unstructured and spatially structured component. Therefore, the covariance matrix of $\xxi$ is
\begin{equation*}
\mbox{Var}({\xxi} | \tau_{\xi}) = \frac{1}{\tau_{\xi}} \left(\phi {\R}_{\ast}^{-} + (1-\phi) {\I}_N \right)
\end{equation*}
expressed as a weighted average of the covariance matrices of the  spatially structured and unstructured components, ${\R}_{\ast}^{-}$ and ${\I}_N$,  respectively. Values of $\phi$ larger than $0.5$ indicate that more than 50\% of the spatial variation is explained by the structured component, indicating the benefits of having a joint model for all regions. For further details we refer to \cite{riebler2016intuitive}.

A RW1 prior distribution is assumed for the temporal random effects ${\ggamma}=(\gamma_1,\ldots,\gamma_T)^{'}$, i.e.
\begin{equation*}
{\ggamma} \sim N({\bf 0}, [\tau_{\gamma} {\R}_{\gamma}]^{-})
\end{equation*}
where $\tau_{\gamma}$ is a precision parameter and ${\R}_{\gamma}$ is the $T \times T$
structure matrix defined as
\begin{equation*}
{\R}_{\gamma} =
\begin{pmatrix}
 1 & -1 & 0 &  &  \\
-1 &  2 & -1 &  &  \\
 & \ddots & \ddots & \ddots &  \\
 &  & -1 & 2 & -1 \\
 &  & 0 & -1 &  1 \\
\end{pmatrix}
\end{equation*}
Finally, for the space-time interaction random effect ${\ddelta}=(\delta_{11},\ldots,\delta_{N1},\ldots,\delta_{1T},\ldots,\delta_{NT})^{'}$, the following prior distribution is assumed
\begin{equation*}
{\ddelta} \sim N({\bf 0},[\tau_{\delta}{\R}_{\delta}]^{-}),
\end{equation*}
where ${\R}_{\delta}$ is the $NT \times NT$ matrix which represents one of the four types of interaction models originally proposed by \cite{knorr2000}. Type I interaction corresponds to a simple adjustment for overdispersion setting with ${\R}_{\delta} = {\I}_T \otimes {\I}_N$, and implies the introduction of iid normally distributed random effects with zero mean and precision $\tau_{\delta}$ for each observation. Type II interaction (${\R}_{\delta} = {\R}_{\gamma} \otimes {\I}_N$) assumes structure in time but not in space, that is, each $\delta_{i\cdot}=(\delta_{i1},\ldots,\delta_{iT})^{'}$ for $i=1,\ldots,N$ follows an independent RW1 prior distribution. Similarly, Type III interaction (${\R}_{\delta} = {\I}_T \otimes {\R}_{\xi}$) assumes structure in space but not in time, that is, each $\delta_{\cdot t}=(\delta_{1t},\ldots,\delta_{Nt})^{'}$ for $t=1,\ldots,T$ follows an intrinsic CAR prior distribution with structure matrix ${\R}_{\xi}$. Finally, for the Type IV interaction, a completely structured precision matrix ${\R}_{\delta} = {\R}_{\gamma} \otimes {\R}_{\xi}$ is assumed.
As for the spatially structured random effect, scaled structure matrices have been considered for the temporal and interaction random effects.

Although these models are flexible enough to describe real situations and their interpretation is fairly straightforward, appropriate sum-to-zero constraints must be imposed on random effects to warrant the identifiability of the intercept, the main spatial and temporal effects, and the space-time interaction effect \citep{schrodle2011,goicoa2018spatio}.

\subsection{Hyperpriors}

In a Bayesian framework, prior distributions need to be assigned to all parameters. Here, we use a uniform improper prior for all standard deviations (square root inverse of precision parameters) in the model \citep[Chapter 5.3]{gelman2006prior, gomez2020bayesian}. A Uniform $[0,1]$ distribution is used for the spatial smoothing parameter $\phi$ of the BYM2 prior. Finally, a vague zero mean normal distribution with a precision close to zero $(0.001)$ is assigned to the intercept $\beta_0$. Additional details on the implementation of these hyperprior distributions in INLA can be found in \cite{ugarte2016}.

\section{Scalable approach for handling large spatial domains}
\label{sec:Scalable_models}
%
When using Model (\ref{eq:M1}) with Type II or Type IV interaction effects, a total of $N-1$ and $N+T-1$ sum-to-zero restrictions on interaction effects are required to avoid identifiability problems, respectively. Consequently, if the number of areas $N$ is very large, the model fitting in  INLA becomes computationally challenging since inference is affected by the number of constraints added to the random effects. Specifically, INLA uses the kriging technique to correct for constraints \citep{rue2005gaussian} with a computational complexity that grows quadratically with the number of constraints. For a high number of constraints, the cost of this technique dominates the overall cost for approximate inference using sparse matrices. 

Recently, \cite{fattah-rue-2022} have proposed a new implementation for fitting this type of spatio-temporal models using INLA based on a dense matrix formulation that automatically imposes the necessary set of identifiability constraints. However, this new approach depends on the accessibility to a high-performance computing architecture to speed-up inference. In addition, a data set with almost $\numprint{8000}$ regions still seems challenging.

Our interest relies on evaluating the scalable Bayesian modelling approach proposed by \cite{orozco2021,orozco2022} for high-dimensional areal count data and extend it for short-term forecasting in time. Using this approach, we divide the spatial domain of interest $\mathscr{D}$  into $D$ subdomains, so that $\mathscr{D} = \bigcup_{d=1}^D \mathscr{D}_d$ where $\mathscr{D}_i \cap \mathscr{D}_j = \emptyset$ for all $i \neq j$. The partition can be chosen based on administrative boundaries, such as states, provinces, autonomous communities or local-health zones, or randomly defined based on a regular grid that is placed over the reference cartography. For our data analyses, partitions are made according to Spanish province boundaries leading to $D=47$ subdomains. Then, separate Bayesian spatio-temporal models are fitted for each subdomain and the results are ultimately joined (denoted as \textit{disjoint model}). One of the main advantages of this scalable approach is that sub-models can be simultaneously fitted using both parallel or distributed computation strategies.
To reduce border effects caused by the partition of the spatial domain, \cite{orozco2021} propose to include $k$-order neighbours for the regions that lie at the boundary of the spatial subdomains. This causes an overlapping of the spatial subdomains, and consequently multiple posterior estimates are obtained for regions lying at the subdomain boundaries. 
Two different merging strategies were compared in \cite{orozco2022} to properly combine the posterior marginal estimates obtained from different sub-models: (i) to weight the estimated posterior probability density functions using mixture distributions, and (ii) to use the posterior marginal distribution given by the model for which the areal-unit of interest originally belongs to. Based on the results obtained from a simulation study, they show that the latter strategy gives better results in terms of risk estimation accuracy and true positive/negative rates. We denote these models as \textit{k-order neighbourhood models}. Furthermore, it was shown that $k=1$ is often a suitable value. The \texttt{R} package \texttt{bigDM} \citep{bigDM} implements this methodology and allows to fit several scalable univariate and multivariate disease mapping models for high-dimensional data in a fully Bayesian setting using INLA.


\subsection{Implementation details}
To fit the classical disease mapping models described in Model (\ref{eq:M1}), calculations are made on a computer with Intel Xeon E5-2620 v4 processors and 256GB RAM (CentOS Linux release 7.3.1611 operative system), using the simplified Laplace approximation strategy in \texttt{R-INLA} (stable version INLA 22.12.16) of R-4.2.0. Partition models, disjoint and first-order neighbourhood models based on $47$ provinces, are distributed over 5 machines (with the same specifications described above) simultaneously running 8 models in parallel on each machine using the \texttt{bigDM} package. The \texttt{R} code to reproduce the results shown in this paper is available at \url{https://github.com/spatialstatisticsupna/Scalable_Prediction}.

\section{Predictive validation study}
\label{sec:ValidationStudy}
To evaluate the predictive ability of all the models, predictions of lung cancer mortality counts in all $\numprint{7907}$ continental municipalities of Spain have been made for $3$ consecutive periods (years). 
We note that mortality registries often provide data with a delay of up to three years.
A total of $K=8$ configurations have been considered, where each configuration uses $T=15$ years of data to fit the model and predict at time points $T+1$, $T+2$ and $T+3$. The first configuration uses data from 1991 to 2005, the second configuration data from 1992 to 2006 and so on. This results in predictions for the years 2006 to 2015, whereby years 2006 and 2015 are only predicted in one configuration, years 2007 and 2014 in two configurations and all the other years in three configurations. \autoref{fig:cascada} of the Appendix~\ref{sec:AppendixB} illustrates the validation setup, which is similar to the one used by \cite{ghosh2007prediction} and \cite{etxeberria2014}.

\subsection{Assessment criteria}
To assess the predictive performance of the models, we compute the mean absolute error (MAE) and the root mean square error (RMSE) of predicted mortality counts for each municipality $i=1,\ldots,7907$ differing between $k={1,2,3}$ year ahead predictions as
\begin{eqnarray*}
	\begin{array}{rcll}
		\mbox{MAE}_i^{(k)} & = & \dfrac{1}{8}\mathlarger{\mathlarger{\sum}}\limits_{t=2005+k}^{2012+k} \left|y_{it} - \widehat{y}_{it} \right|,
		\\[5.ex]
		\mbox{RMSE}_i^{(k)} & = & \sqrt{\dfrac{1}{8} \mathlarger{\mathlarger{\sum}}\limits_{t=2005+k}^{2012+k} \left( y_{it} - \widehat{y}_{it} \right)^2},
	\end{array}
\end{eqnarray*}
where $y_{it}$ are the number of observed cases and $\widehat{y}_{it}$ is the expected value for the posterior predictive counts for areal unit $i$ and time period $t$, respectively. See Appendix~\ref{sec:AppendixA} for details about its computation with INLA. Looking at \autoref{fig:cascada} this means that average scores over the orange, yellow and blue cells are built.

To assess not only point predictions but the entire predictive distribution, we compute the 95\% interval score (IS), which is a proper scoring rule that combines calibration and sharpness of predictions \citep{gneiting2007strictly}. This measure transforms interval width and empirical coverage into a single score, and has recently become popular (see for example \citealp{hofer2022,paige2022}). Let $y$ be the observed number of cases and  $[l,u]$ be the respective $(1-\alpha)\cdot 100\%$ posterior predictive credible interval at credible level $\alpha \in (0,1)$, then
\begin{eqnarray*}
	IS_{\alpha}(y) = (u-l)+\dfrac{2}{\alpha}(l-y)I[y<l] +\dfrac{2}{\alpha}(y-u)I[y>u].
\end{eqnarray*}
Here, $I[\cdot]$ denotes an indicator function that penalises the length of the credible interval if the number of observed cases is not contained within that interval.

\subsection{Results}

\autoref{tab:ValidationStudy} compares average values of 95\% interval score (IS), MAE and RMSE over all the municipalities for 1, 2 and 3-years ahead predictions when using the different models described in the previous section. For all these criteria, lower values are preferred. As expected, higher scores are obtained as we increase the time of prediction in future years. Note that we were not able to fit with INLA the classical disease mapping models using Type II and Type IV interactions due to the high-number of identifiability constraints. However, we were able to fit the partition models proposed by \cite{orozco2022} with all types of space-time interactions. In addition, a substantial reduction in computational time for fitting Type I and Type III models was obtained.
We observe that partition models outperform the classical models when these latter models can be fitted (only Type I and Type III interactions). The best predictive measures are obtained using the Type IV interaction model. Compared to a disjoint model that estimates all $47$ provinces separately we find that the 1st-order neighbourhood models obtains slightly better IS values when predicting 2 and 3-years ahead. The running time is slightly more than twice as long.

\begin{table}[!ht]
\caption{Validation study: average values of evaluation scores (interval score -IS-, mean absolute error -MAE- and root mean square error- RMSE-) and computational time (in minutes) for classical, disjoint and 1st-order neighbourhood models fitted using the simplified Laplace approximation strategy of INLA.}
\label{tab:ValidationStudy}
\vspace{-0.5cm}
\renewcommand{\arraystretch}{1.1}
\begin{center}
\resizebox{\textwidth}{!}{
\begin{tabular}{llrrrrrrrrrrrr}
\hline
\multirow{2}{20mm}{Model} & \multirow{2}{20mm}{Space-time interaction} & \multicolumn{3}{c}{1-year ahead} & & \multicolumn{3}{c}{2-years ahead} & &\multicolumn{3}{c}{3-years ahead} & \multirow{2}{*}{Time} \\
\cline{3-5} \cline{7-9} \cline{11-13}
& & IS$_{0.05}$ & MAE & RMSE & & IS$_{0.05}$ & MAE & RMSE & & IS$_{0.05}$ & MAE & RMSE & \\
\hline
Classical     & Type I   & 3.91 & 0.68 & 0.84 & & 3.97 & 0.69 & 0.85 & & 3.97 & 0.70 & 0.85 &  343\\
              & Type II  &  $-$ &  $-$ &  $-$ & &  $-$ &  $-$ &  $-$ & &  $-$ &  $-$ &  $-$ &  $-$\\
              & Type III & 3.94 & 0.68 & 0.84 & & 3.96 & 0.69 & 0.85 & & 4.03 & 0.70 & 0.86 & 1767\\
              & Type IV  &  $-$ &  $-$ &  $-$ & &  $-$ &  $-$ &  $-$ & &  $-$ &  $-$ &  $-$ &  $-$\\
\hline
Disjoint      & Type I   & 3.91 & 0.68 & 0.83 & & 3.95 & 0.69 & 0.84 & & 3.97 & 0.70 & 0.85 &   14\\
by provinces  & Type II  & 3.87 & 0.67 & 0.82 & & 3.92 & 0.68 & 0.83 & & 3.98 & 0.69 & 0.84 &  152\\
              & Type III & 3.88 & 0.68 & 0.83 & & 3.92 & 0.69 & 0.84 & & 3.95 & 0.69 & 0.85 &   22\\
              & Type IV  & \textbf{3.84} & \textbf{0.67} & \textbf{0.82} & & \textbf{3.90} & \textbf{0.68} & \textbf{0.83} & & \textbf{3.96} & \textbf{0.69} & \textbf{0.84} &  \textbf{179}\\
\hline
1st order     & Type I   & 3.90 & 0.68 & 0.83 & & 3.94 & 0.69 & 0.84 & & 3.96 & 0.69 & 0.85 &   18\\
neighbourhood & Type II  & 3.86 & 0.67 & 0.82 & & 3.91 & 0.68 & 0.83 & & 3.97 & 0.69 & 0.84 &  405\\
by provinces  & Type III & 3.88 & 0.68 & 0.83 & & 3.91 & 0.69 & 0.84 & & 3.93 & 0.69 & 0.85 &   40\\
             & Type IV  & \textbf{3.84} & \textbf{0.67} & \textbf{0.82} & & \textbf{3.88} & \textbf{0.68} & \textbf{0.83} & & \textbf{3.93} & \textbf{0.69} & \textbf{0.84} &  \textbf{433}\\
\hline
\end{tabular}}
\end{center}
\end{table}

\begin{figure}[!ht]
    \centering
    \includegraphics[width=0.95\textwidth]{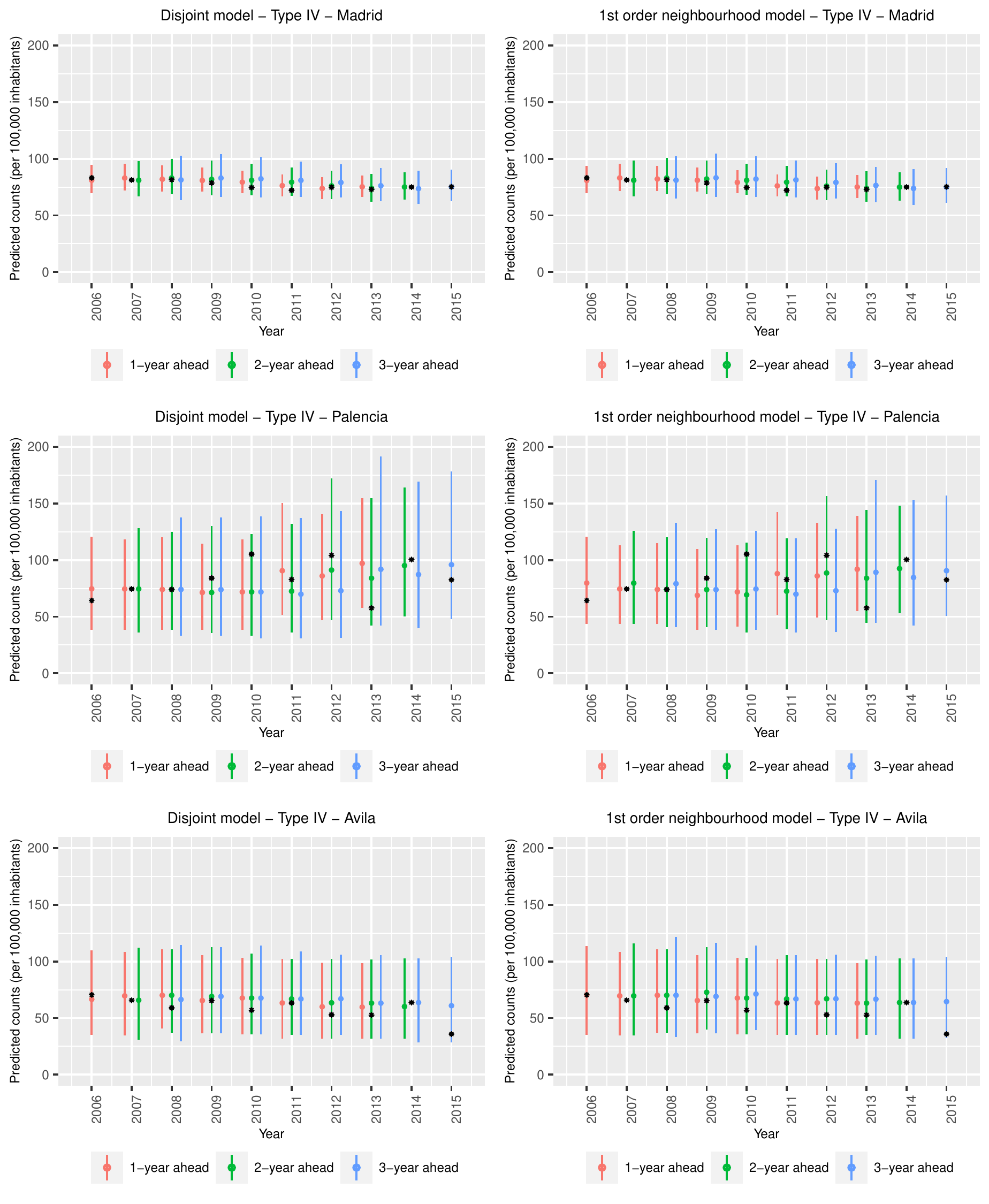}
    \caption{One, two and three-year ahead predictions for the municipalities of Madrid (top), Palencia (middle) and {\'A}vila (bottom) using the disjoint model (left column) and 1st-order neighbourhood model (right column) with Type IV interactions. Expected values of posterior predictive counts per \numprint{100000} inhabitants (dots) and 95\% predictive intervals (color lines) are plotted. The number of real observed number of cases are also included as black stars.}
    \label{fig:ValidationStudy}
\end{figure}

\autoref{fig:ValidationStudy} investigates predictions in more detail for three selected provincial capitals: the municipalities of Madrid, Palencia and {\'A}vila. Here, expected values for the posterior predictive counts (deaths per $\numprint{100000}$ inhabitants) together with 95\% predictive intervals are plotted using the disjoint and 1st-order neighbourhood models with Type IV interaction. Different colours are used for 1, 2 and 3-years ahead predictions.
Very similar forecasts are obtained for the municipality of Madrid (a region with a large number of population at risk) under the different models. As expected, wider predictive intervals are obtained for those areas with lower values of observed cases and population at risk, as is the case of the municipalities of Palencia and {\'A}vila. For these municipalities, slightly wider predictive intervals are observed when using the disjoint model.

To perform a more in-depth analysis between the predictive performance of the partition models, we compute average values of prediction evaluation scores for two subsets of the data: (i) municipalities for which the proportion of zero observed cases during the study period is less or equal than 0.2, and (ii) municipalities lying at the boundary between two or more provinces with at least 2 observed cases per $\numprint{100000}$ inhabitants during the whole study period. The results are shown in \autoref{tab:ValidationStudy_PropZero} and \autoref{tab:ValidationStudy_Boundary}. In both scenarios, the 1st-order neighbourhood model outperforms the disjoint model in terms of prediction accuracy and interval score.

\section{Illustration: projections of cancer mortality in Spain}
\label{sec:DataAnalysis}

The aim of this section is to illustrate our proposal for forecasting cancer burden up to three years, as cancer registries often suffer from this delay in data provision. Here we use both male lung cancer and overall cancer (all sites) mortality data in the $\numprint{7907}$ municipalities of continental Spain in the period 1991-2012 to forecast cancer data for the years 2013, 2014, 2015.  This allows us to check whether actual rates are close to predicted rates. The same models described in the validation study of Section~\ref{sec:ValidationStudy} have been considered here.

To compare the predictive performance of the different models we use cross-validation techniques to compute scoring rules based on the estimated predictive distribution of the mortality counts. Typically, cross-validation techniques are based on the idea of splitting the observed data into a training set (sample data used for model's parameter estimation) and a testing set (set of points used to compute the prediction error based on the training model) multiple times to estimate the predictive accuracy of the model \citep{gelman1995bayesian,hastie2009elements}.
A particular interesting feature of INLA is that it provides leave-one-out cross-validatory (LOOCV) model checks without re-running the model for each observation in turn \citep{rue2009approximate,held2010posterior}, something that would be computationally unfeasible when fitting complex models on large data sets. Specifically, INLA provides approximations of the conditional predictive ordinates (CPO, \citealp{pettit1990}), $\mbox{CPO}_{it}=\pi(Y_{it} = y_{it} | {\yy}_{-it})$, which is defined as the cross-validates predictive probability mass at the observed count $y_{it}$.
However, it is well-known that LOOCV techniques may not be appropriate to measure the predictive performance of a model that includes spatially and/or temporally structured random effects to deal with correlated data \citep{roberts2017cross,rabinowicz2022cross}. To solve this problem, \cite{liu2022leave} propose an automatic group construction procedure for leave-group-out cross-validation (LGOCV) to estimate the predictive performance of structured models for latent Gaussian models with INLA.

\clearpage

The groups derived from the automatic construction when fitting the 1st-order neighbourhood model is used as a reference for the rest of the models to make predictive measures comparable. \autoref{tab:LS} shows the sum of the log-predictive densities computed over each area-time point using both the LOOCV (usually named as logarithmic score \citealp{gneiting2007strictly}) and the LGOCV techniques, as well as model selection criteria such as the deviance information criterion (DIC; \citealp{Spiegelhalter2002}) and the Watanabe-Akaike information criterion (WAIC; \citealp{watanabe2010asymptotic}). For comparison purposes, the reference value has been set to zero by subtracting the minimum value for each column when computing the cross-validation measures.
As in the validation study, we were not able to fit the classical spatio-temporal models with Type II and Type IV interactions. For both lung and overall cancer mortality data analyses, partition models show better predictive performance and better values for model selection criteria (see \autoref{tab:LS}). In particular, the 1st-order neighbourhood models with Type IV interactions. As expected, the differences are more pronounced when comparing the predictive performance of the models for overall cancer data, as a higher number of deaths are observed for each areal time unit compared to lung cancer data.

\begin{table}[ht]
\caption{Logarithmic score using both LOOCV and LGOCV techniques, model selection criteria and computational time (in minutes) for lung cancer mortality and overall cancer mortality data with models fitted using the simplified Laplace approximation strategy of INLA (stable version INLA\_22.12.16).}
\label{tab:LS}
\vspace{-0.5cm}
\renewcommand{\arraystretch}{1.2}
\begin{center}
\resizebox{\textwidth}{!}{
\begin{tabular}{llrrllrrrrllr}
\hline
& & \multicolumn{3}{c}{Lung cancer} & & & & \multicolumn{3}{c}{Overall cancer} & & \\
\cline{1-7} \cline{9-13}
& & LOOCV* & LGOCV* & DIC & WAIC & T. total & & LOOCV* & LGOCV* & DIC & WAIC & T. total \\
\hline
Classical     & Type I   & 440 & 478 & 130383 & 130480 &  622 & & 1542 & 1680 & 182280 & 182613 &  598 \\
              & Type II  & $-$ & $-$ &    $-$ &    $-$ &  $-$ & &  $-$ & $-$  &    $-$ &    $-$ &  $-$ \\
              & Type III & 365 & 388 & 130241 & 130373 & 3914 & & 1110 & 1110 & 181470 & 181856 & 4674 \\
              & Type IV  & $-$ & $-$ &    $-$ &    $-$ &  $-$ & &  $-$ & $-$  &    $-$ &    $-$ &  $-$ \\
\cline{1-13}
Disjoint      & Type I   & 244 & 261 & 130122 & 130150 &  20 & &  925 &  975 & 181206 & 181393 &  20  \\
by provinces  & Type II  &  52 &  81 & 129835 & 129908 & 296 & &   46 &  145 & 179813 & 180018 & 275 \\
              & Type III & 209 & 212 & 130076 & 130147 &  36 & &  836 &  845 & 181075 & 181361 &  36  \\
              & Type IV  &   0 &   8 & 129733 & 129824 & 379 & &   18 &   49 & 179766 & 180024 & 323 \\
\cline{1-13}
1st order     & Type I   & 234 & 247 & 130061 & 130110 &  28 & &  912 &  936 & 181141 & 181338 &  28  \\
neighbourhood & Type II  &  47 &  69 & 129778 & 129856 & 661 & &   57 &  129 & 179803 & 180012 & 633 \\
by provinces  & Type III & 209 & 213 & 130035 & 130128 &  55 & &  799 &  786 & 180991 & 181293 &  55  \\
              & Type IV  &   \textbf{0} &   \textbf{0} & \textbf{129685} & \textbf{129777 }& 800 & &    \textbf{0} &    \textbf{0} & \textbf{179690} & \textbf{179951} & 754 \\
\hline
\multicolumn{9}{l}{$\ast$Note: Reference value has been set to zero by subtracting the minimum value for each column.}& & & &
\end{tabular}}
\end{center}
\end{table}

\subsection{Lung cancer mortality}\label{sec:LC mortality}
In this section, we provide lung cancer mortality projections in the municipalities of Spain for the period 2013-2015 using the 1st-order neighbourhood model with Type IV space-time interactions. \autoref{tab:LungCancer_projections} shows posterior median estimates of the predicted mortality rates per $\numprint{100000}$ males and its corresponding 95\% credible intervals for years 2013 and 2015 for the 47 municipalities that are provincial capitals. As expected, wider credible intervals are obtained when computing 3-year ahead predictions (year 2015).

\begin{table}[ht]
\caption{Posterior median estimates of the lung cancer mortality rates per $\numprint{100000}$ males and its corresponding 95\% credible intervals for years 2013 and 2015 for the 47 municipalities that form the provincial capitals.}
\label{tab:LungCancer_projections}
\vspace{-0.5cm}
\renewcommand{\arraystretch}{1.1}
\begin{center}
\resizebox{\textwidth}{!}{
\begin{tabular}{llllll|llllll}
\hline
& \multicolumn{2}{c}{2013} & &\multicolumn{2}{c}{2015} &  & \multicolumn{2}{c}{2013} & & \multicolumn{2}{c}{2015}\\
\cline{1-3} \cline{5-9} \cline{11-12}
Municipality  & $\hat{\lambda}_{it^{\star}}\cdot 10^5$ & 95\% CI & & $\hat{\lambda}_{it^{\star}}\cdot 10^5$& 95\% CI & Municipality  & $\hat{\lambda}_{it}\cdot 10^5$ & 95\% CI & & $\hat{\lambda}_{it}\cdot 10^5$ & 95\% CI\\
\hline
Guadalajara       &  56.1 &  (35.9,85.2) & & 57.2 & (34.3,87.0)  & Pontevedra        &  78.8 & (48.3,111.8) & &  76.8 & (48.6,115.2)\\
Ja{\'e}n          &  58.8 &  (37.4,82.0) & & 59.4 & (37.8,84.6)  & Ciudad Real       &  79.0 & (48.0,115.6) & &  76.7 & (48.3,119.3)\\
Albacete          &  59.2 &  (41.8,80.1) & & 59.6 & (40.9,81.7)  & Huesca            &  83.2 & (47.6,126.8) & &  84.1 & (48.0,128.1)\\
Girona            &  59.9 &  (38.5,87.7) & & 59.9 & (36.4,89.8)  & Valladolid        &  84.0 & (66.2,104.5) & &  84.5 & (62.9,109.6)\\
Segovia           &  61.6 & (30.8,104.0) & & 64.0 & (31.9,108.0) & Huelva            &  84.0 & (61.6,112.0) & &  83.9 & (58.3,115.2)\\
{\'A}vila         &  63.2 & (35.1,101.8) & & 64.5 & (32.3,107.5) & Badajoz           &  84.3 & (62.3,109.0) & &  83.7 & (60.1,112.4)\\
Murcia            &  63.3 &  (51.2,76.3) & & 63.2 & (50.2,78.1)  & Pamplona          &  85.1 & (64.9,107.5) & &  85.9 & (64.4,110.6)\\
Soria             &  63.6 & (26.5,111.3) & & 59.3 & (21.6,113.2) & C{\'a}ceres       &  86.7 & (60.7,119.3) & &  89.2 & (56.6,124.1)\\
Granada           &  64.2 &  (47.0,84.1) & & 64.9 & (44.8,88.7)  & San Sebasti{\'a}n &  86.8 & (62.8,113.1) & &  86.8 & (59.4,118.8)\\
Tarragona         &  65.9 &  (45.4,89.3) & & 65.9 & (44.9,91.3)  & Valencia          &  87.4 & (75.3,100.9) & &  87.4 & (73.3,103.1)\\
Burgos            &  66.5 &  (46.6,86.3) & & 66.3 & (45.0,91.1)  & Barcelona         &  88.1 &  (77.8,99.1) & &  88.2 & (74.5,103.2)\\
Castell{\'o}n     &  66.7 &  (48.6,89.4) & & 67.1 & (48.0,89.9)  & Zamora            &  88.4 & (52.4,134.2) & &  86.9 & (46.8,140.4)\\
Toledo            &  67.5 &  (42.5,97.5) & & 67.7 & (40.1,100.4) & Palencia          &  89.3 & (55.1,133.9) & &  90.6 & (50.6,146.6)\\
Vitoria           &  69.2 &  (52.3,89.5) & & 68.7 & (49.4,93.0)  & Zaragoza          &  90.9 & (76.4,108.1) & &  91.2 & (70.7,113.6)\\
Almer{\'i}a       &  69.3 &  (50.1,91.8) & & 68.8 & (47.6,94.2)  & Ourense           &  91.2 & (60.8,125.6) & &  90.5 & (55.6,133.8)\\
Logro{\~n}o       &  69.5 &  (49.1,96.8) & & 69.4 & (47.2,101.3) & Lugo              &  93.3 & (62.9,128.0) & &  94.2 & (61.4,133.7)\\
L{\'e}rida        &  72.2 &  (50.5,98.1) & & 71.6 & (48.2,102.3) & Bilbao            &  95.8 & (75.2,118.8) & &  95.3 & (70.7,124.3)\\
C{\'o}rdoba       &  74.7 &  (59.5,91.2) & & 75.0 & (59.1,92.8)  & Santander         &  97.6 & (72.0,128.1) & &  98.4 & (67.2,135.7)\\
Teruel            &  74.9 & (40.2,126.7) & & 76.2 & (35.2,129.0) & A Coru{\~n}a          &  98.9 & (77.0,126.0) & &  98.9 & (73.3,133.4)\\
Madrid            &  74.9 &  (65.5,85.0) & & 75.0 &  (61.1,91.1) & Salamanca         &  99.0 & (69.9,135.4) & &  98.7 & (62.8,148.1)\\
M{\'a}laga        &  75.0 &  (62.5,88.1) & & 74.9 & (60.6,90.2)  & Le{\'o}n          & 102.2 & (73.7,139.0) & & 103.3 & (68.8,144.6)\\
Cuenca            &  77.9 & (40.8,118.7) & & 75.5 & (41.5,120.7) & Oviedo            & 103.9 & (81.0,131.6) & & 104.8 & (75.7,138.8)\\
Sevilla           &  78.4 &  (67.3,90.8) & & 78.6 &  (65.2,93.4) & C{\'a}diz         & 113.3 & (80.7,154.6) & & 114.0 & (73.7,168.4)\\
Alicante          &  78.6 &  (64.5,94.6) & & 79.0 &  (64.0,94.7) & & & & & & \\
\hline
\end{tabular}}
\end{center}
\end{table}

\begin{figure}[!ht]
    \centering
    \vspace{-1cm}
    \includegraphics[width=1.25\textwidth]{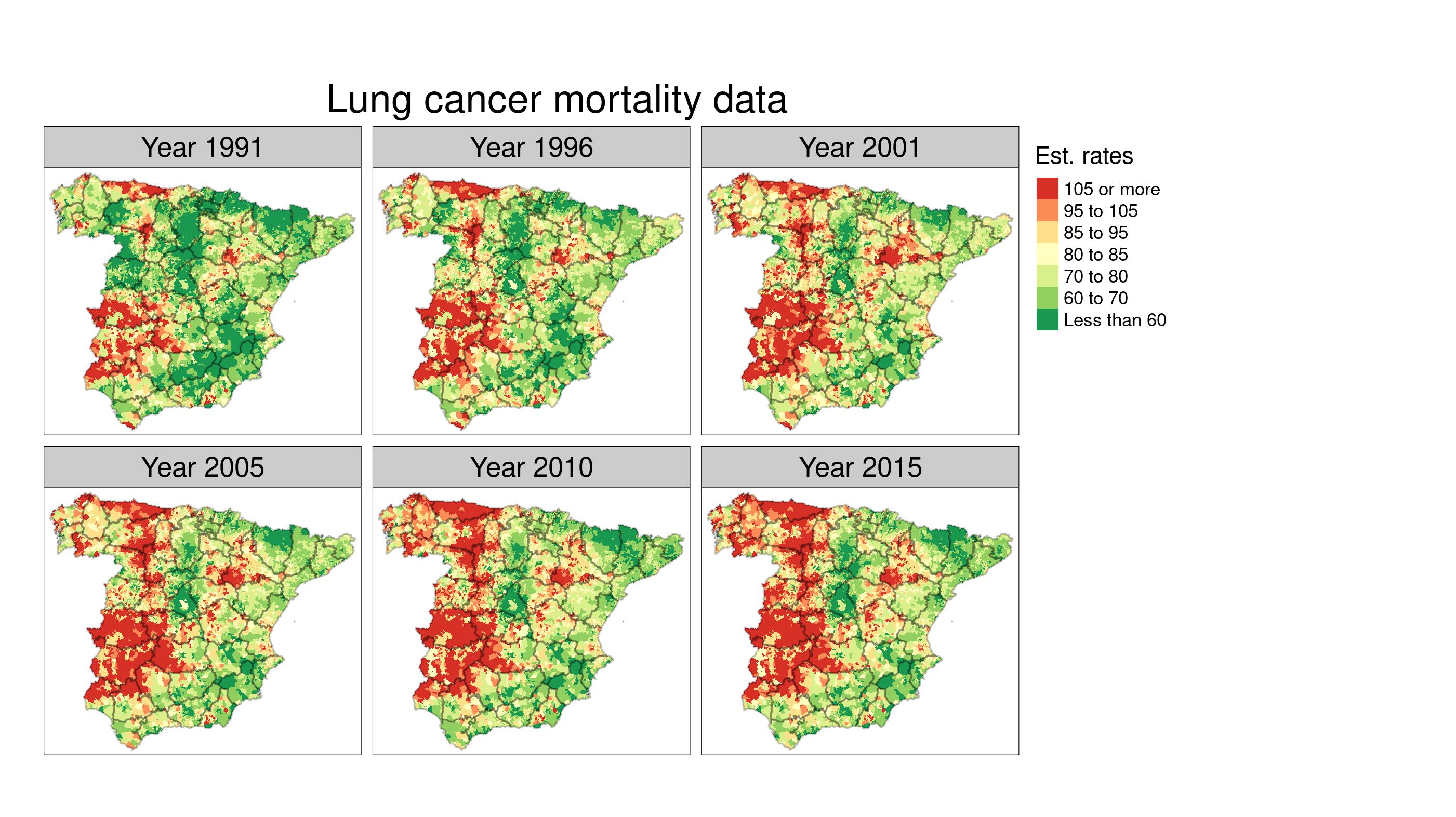}\\[-0.75cm]
    \includegraphics[width=1.25\textwidth]{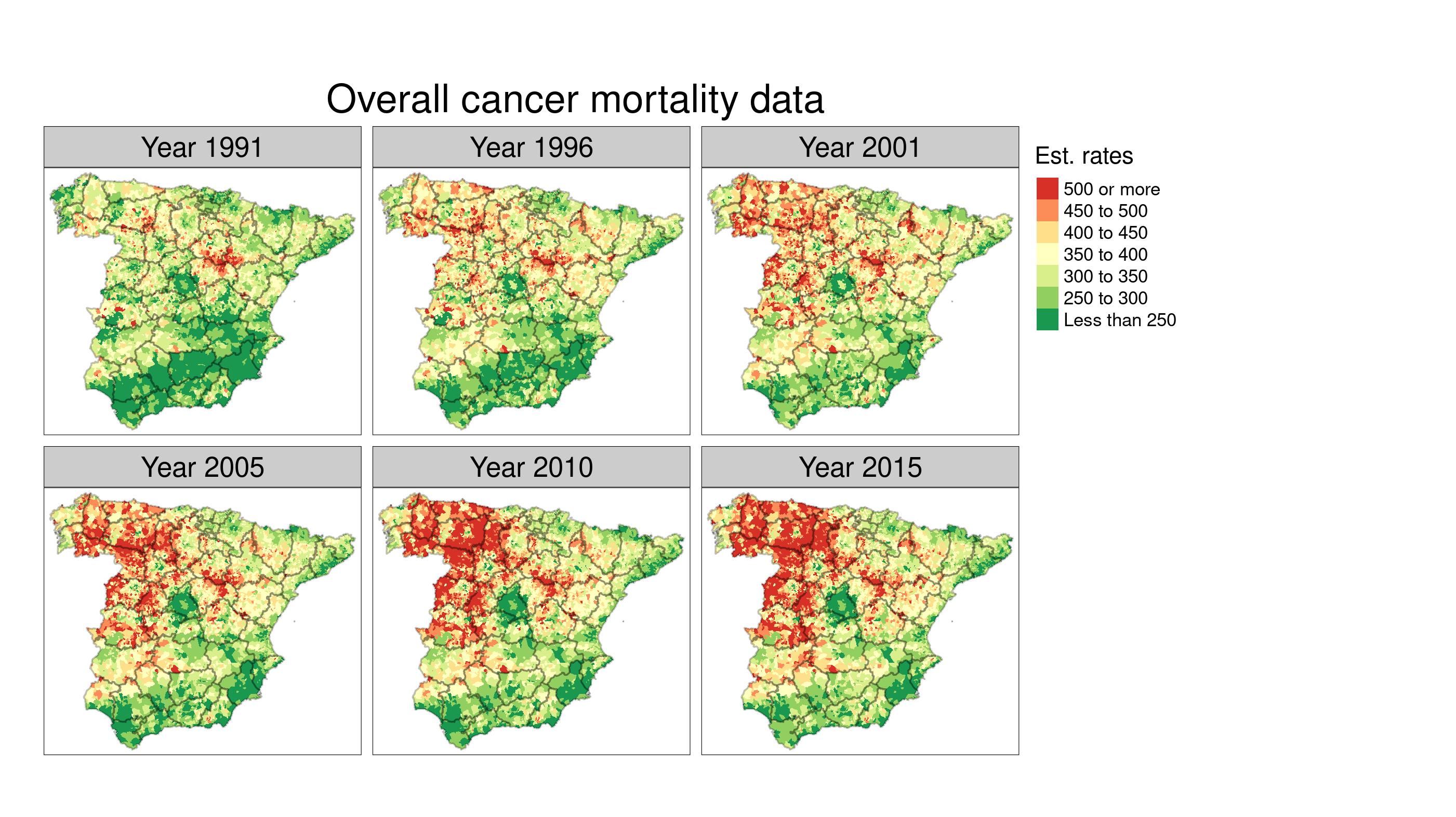}\\
    \vspace{-0.75cm}
    \caption{Posterior median estimates of lung cancer (top) and overall cancer (bottom) mortality rates per $\numprint{100000}$ males for the $\numprint{7907}$ municipalities of continental during the period 1991-2015. Years 2013 to 2015 were predicted.}
    \label{fig:Map_EstRates}
\end{figure}

\autoref{fig:Map_EstRates} (top) shows the maps with the temporal evolution of posterior median estimates of lung cancer mortality rates for some selected years between 1991 and 2015. 
An increasing trend is observed in the North-West and Central-West regions of Spain, in particular during the period 1991-2001. The average mortality rate for Spain is about 71.7 deaths per $\numprint{100000}$ males in the year 1991, rising to 78.3 and 83.2 deaths per $\numprint{100000}$ males in the years 1996 and 2001, respectively. For the second half of the period, the estimated rates are fairly constant without major spatial changes (average mortality rates close to 83.5 deaths per $\numprint{100000}$ males during the years 2005-2010, with a slight increase to 84.5 deaths per $\numprint{100000}$ males in 2012), something that is also observed for the predicted years 2013 to 2015.

\begin{figure}[!ht]
    \centering
    \includegraphics[scale=0.85]{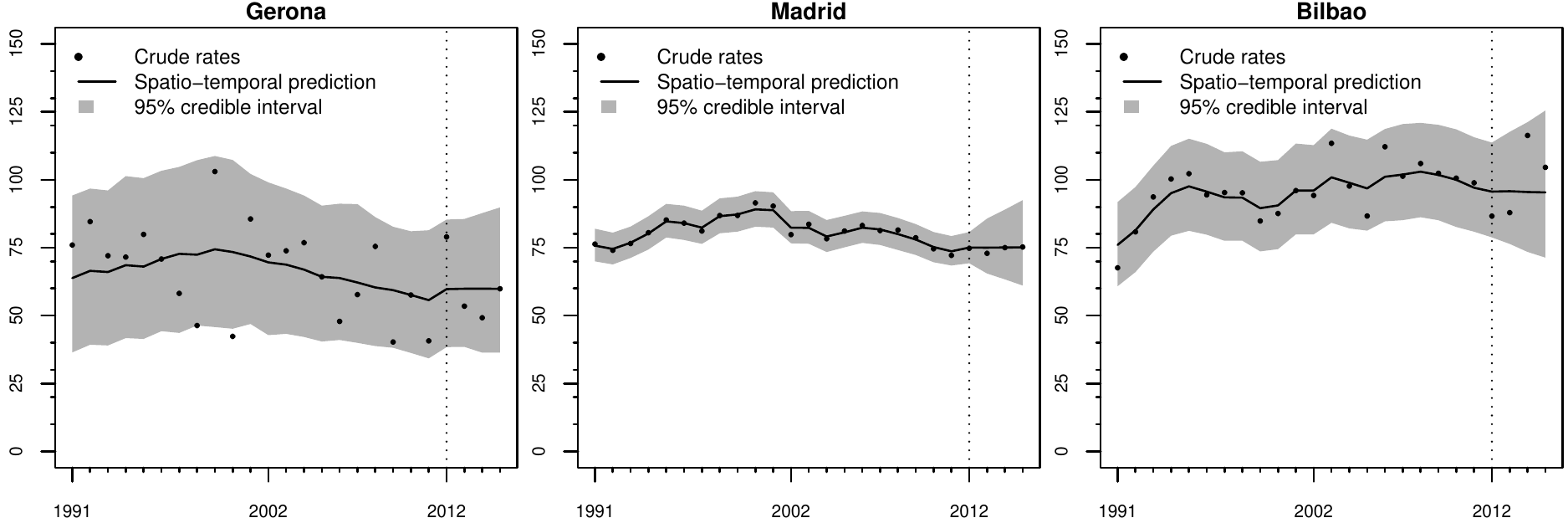}\\[0.5cm]
    \includegraphics[scale=0.85]{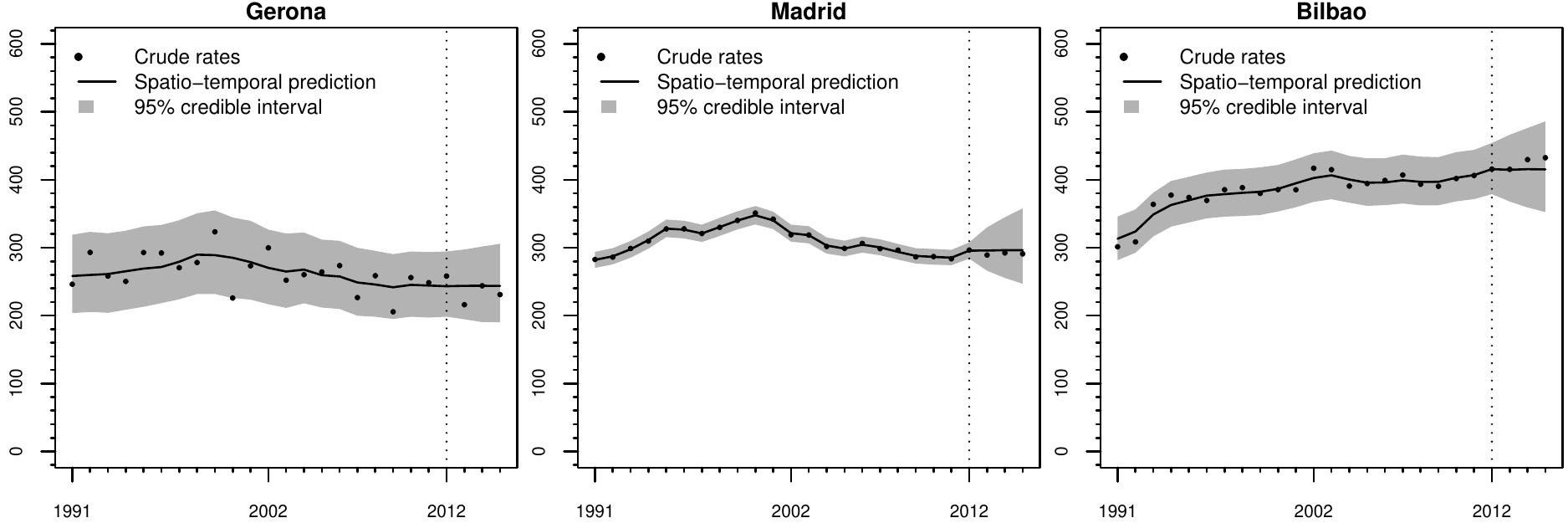}\\
    \caption{Posterior predictive median and its corresponding 95\% credible interval of lung cancer (top) and overall cancer (bottom) mortality rates per \numprint{100000} males during the period 1991-2015 for the municipalities of Girona, Madrid, and Bilbao. Crude rates are shown as a filled circle. The vertical dotted line indicates where the prediction starts.}
    \label{fig:Trends_Muni}
\end{figure}

In \autoref{fig:Trends_Muni} (top) we show the temporal evolution of mortality rate forecasts for the provincial capitals of Girona, Madrid and Bilbao (selected to show areas with different estimated temporal trends). In general, the 95\% credible intervals contain the crude rates over all the study period. As expected, wider credible intervals are obtained for those areas with lower population at risk.

\clearpage


\autoref{fig:Percentage_change} (top) shows the posterior median of the percentage change in lung cancer mortality rates between 2013 and 2015 at municipality level (see Appendix~\ref{sec:AppendixA} for computational details). That is, the posterior median of $100\% \times [(\lambda_{i\text{2015}} -\lambda_{i\text{2013}})/\lambda_{i\text{2013}}]$. Positive values indicate a growth in the 2015 forecast compared to 2013. 
Variations between $-0.20\%$ and $0.04\%$ are the most frequent in all municipalities, i.e.,  quite close to zero. This result shows that the predictions in the municipalities of continental Spain have slight variations as it can be seen in  \autoref{fig:Map_EstRates}.

\subsection{Overall cancer mortality}
Similar to the previous section, here we provide overall cancer mortality projections in the municipalities of Spain for the period 2013-2015 using the 1st-order neighbourhood model with Type IV space-time interactions. \autoref{tab:OverallCancer_projections} provides posterior median estimates of predicted mortality rates per $\numprint{100000}$ males and its corresponding 95\% credible intervals for the provincial capitals. The municipalities of Santander, Bilbao and Salamanca have the highest overall cancer mortality rates, while Murcia, Toledo and Guadalajara shows the lowest values. In general, the temporal trend in forecasts is quite stable with slight variation in some areas, as is the case of Soria with an increase of 5 cases per $\numprint{100000}$ males.

\begin{table}[!ht]
\caption{Posterior median estimates of the predicted overall cancer mortality rates per $\numprint{100000}$ males and its corresponding 95\% credible intervals for years 2013 and 2015 for the 47 municipalities that form the provincial capitals.}
\label{tab:OverallCancer_projections}
\vspace{-0.5cm}
\renewcommand{\arraystretch}{1.1}
\begin{center}
\resizebox{\textwidth}{!}{
\begin{tabular}{llllll|llllll}
\hline
& \multicolumn{2}{c}{2013} & &\multicolumn{2}{c}{2015} &  & \multicolumn{2}{c}{2013} & & \multicolumn{2}{c}{2015}\\
\cline{1-3} \cline{5-9} \cline{11-12}
Municipality  & $\hat{\lambda}_{it^{\star}}\cdot 10^5$ & 95\% CI & & $\hat{\lambda}_{it^{\star}}\cdot 10^5$& 95\% CI & Municipality  & $\hat{\lambda}_{it}\cdot 10^5$ & 95\% CI & & $\hat{\lambda}_{it}\cdot 10^5$ & 95\% CI\\
\hline
Murcia            & 228.9 & (204.3,255.5) & & 229.6 & (199.4,262.1) & Vitoria           & 311.5 & (267.6,357.1) & & 310.8 & (261.4,368.6)\\
Guadalajara       & 239.9 & (188.4,298.2) & & 240.3 & (180.8,309.0) & Valencia          & 316.9 & (288.1,347.2) & & 316.8 & (279.8,362.4)\\
Toledo            & 242.4 & (184.9,304.9) & & 243.4 & (180.6,313.6) & Pamplona          & 322.4 & (279.8,369.2) & & 322.1 & (275.9,375.8)\\
Albacete          & 242.6 & (204.3,285.6) & & 242.9 & (199.7,294.2) & Zaragoza          & 323.1 & (290.5,357.9) & & 324.4 & (278.6,378.0)\\
Girona            & 243.9 & (194.7,297.4) & & 243.8 & (192.5,307.9) & Pontevedra        & 327.9 & (266.9,396.5) & & 327.6 & (258.5,404.4)\\
Castell{\'o}n     & 248.9 & (204.7,298.6) & & 249.4 & (195.4,311.7) & Soria             & 328.6 & (243.8,429.3) & & 334.1 & (247.9,431.1)\\
Ja{\'e}n          & 253.1 & (203.2,310.1) & & 253.8 & (190.8,325.8) & Barcelona         & 335.7 & (305.9,369.5) & & 335.9 & (289.1,386.3)\\
Almer{\'i}a       & 255.0 & (214.4,298.7) & & 254.0 & (207.5,305.9) & Teruel            & 357.2 & (259.2,472.4) & & 357.7 & (252.1,492.6)\\
Tarragona         & 256.2 & (213.8,303.1) & & 255.9 & (209.5,308.3) & Burgos            & 367.4 & (317.2,424.5) & & 370.5 & (305.4,439.2)\\
M{\'a}laga        & 266.6 & (236.6,297.7) & & 267.3 & (229.0,310.8) & A Coru{\~n}a      & 368.4 & (313.3,428.8) & & 368.4 & (296.0,457.6)\\
L{\'e}rida        & 277.1 & (226.6,331.9) & & 276.3 & (219.3,345.0) & San Sebasti{\'a}n & 368.9 & (317.5,422.6) & & 367.9 & (308.5,433.0)\\
C{\'o}rdoba       & 277.9 & (244.4,313.4) & & 277.2 & (238.4,323.6) & Valladolid        & 370.8 & (320.9,427.4) & & 372.2 & (303.1,451.8)\\
Alicante          & 280.6 & (244.4,323.0) & & 279.8 & (231.5,338.8) & Huesca            & 372.6 & (285.4,467.8) & & 372.3 & (276.2,492.3)\\
Granada           & 282.0 & (238.6,329.0) & & 281.6 & (228.6,345.6) & Zamora            & 376.4 & (288.1,481.2) & & 377.8 & (267.5,511.7)\\
Ciudad Real       & 282.1 & (222.8,352.6) & & 284.1 & (215.9,363.6) & Lugo              & 381.8 & (314.6,462.1) & & 383.5 & (302.4,479.9)\\
Badajoz           & 288.0 & (242.6,336.0) & & 286.3 & (234.0,349.0) & Palencia          & 396.5 & (320.3,483.1) & & 397.2 & (306.5,501.2)\\
Cuenca            & 289.4 & (218.9,371.0) & & 290.5 & (211.3,384.8) & C{\'a}diz         & 398.4 & (310.8,498.0) & & 398.2 & (282.4,545.6)\\
Sevilla           & 291.5 & (265.4,320.7) & & 291.6 & (257.6,330.4) & Oviedo            & 403.2 & (351.8,458.5) & & 402.7 & (338.6,476.4)\\
C{\'a}ceres       & 292.7 & (238.5,355.7) & & 293.8 & (230.7,367.8) & Le{\'o}n          & 405.3 & (331.6,489.0) & & 406.2 & (309.8,519.8)\\
{\'A}vila         & 294.9 & (221.2,382.7) & & 293.9 & (211.5,394.2) & Ourense           & 407.2 & (336.3,486.2) & & 407.5 & (323.1,504.2)\\
Madrid            & 295.8 & (265.3,329.1) & & 295.5 & (248.0,353.5) & Salamanca         & 412.0 & (345.0,492.1) & & 412.8 & (324.5,523.5)\\
Segovia           & 300.4 & (223.4,389.0) & & 300.0 & (212.0,411.9) & Bilbao            & 414.7 & (368.6,465.6) & & 414.6 & (355.6,484.1)\\
Huelva            & 302.5 & (246.5,361.3) & & 303.1 & (234.8,381.3) & Santander         & 419.8 & (338.0,517.4) & & 422.1 & (306.3,571.5)\\
Logro{\~n}o       & 306.6 & (254.8,365.2) & & 306.6 & (245.6,384.3) & & & & & & \\
\hline\\[-1cm]
\end{tabular}}
\end{center}
\end{table}

The maps with the temporal evolution of the overall cancer mortality rates for the $\numprint{7907}$ municipalities of continental Spain are shown at the bottom of \autoref{fig:Map_EstRates}. We observe a remarkable increase in the estimated cancer rates in the regions located in the west (Extremadura) and northwest (Castilla y Le{\'o}n, Galicia, Asturias, and part of Cantabria) during the period 1991-2015. 
The overall mortality rate for the whole of Spain shows a steady rise during the study period, with average values of  325.5, 355.5, 373.4, 381.6 and 386.3 deaths per $\numprint{100000}$ males in the years 1991, 1996, 2001, 2010 and 2015, respectively.

\begin{figure}[ht]
    \centering
    \vspace{-1cm}
    \includegraphics[scale=0.65]{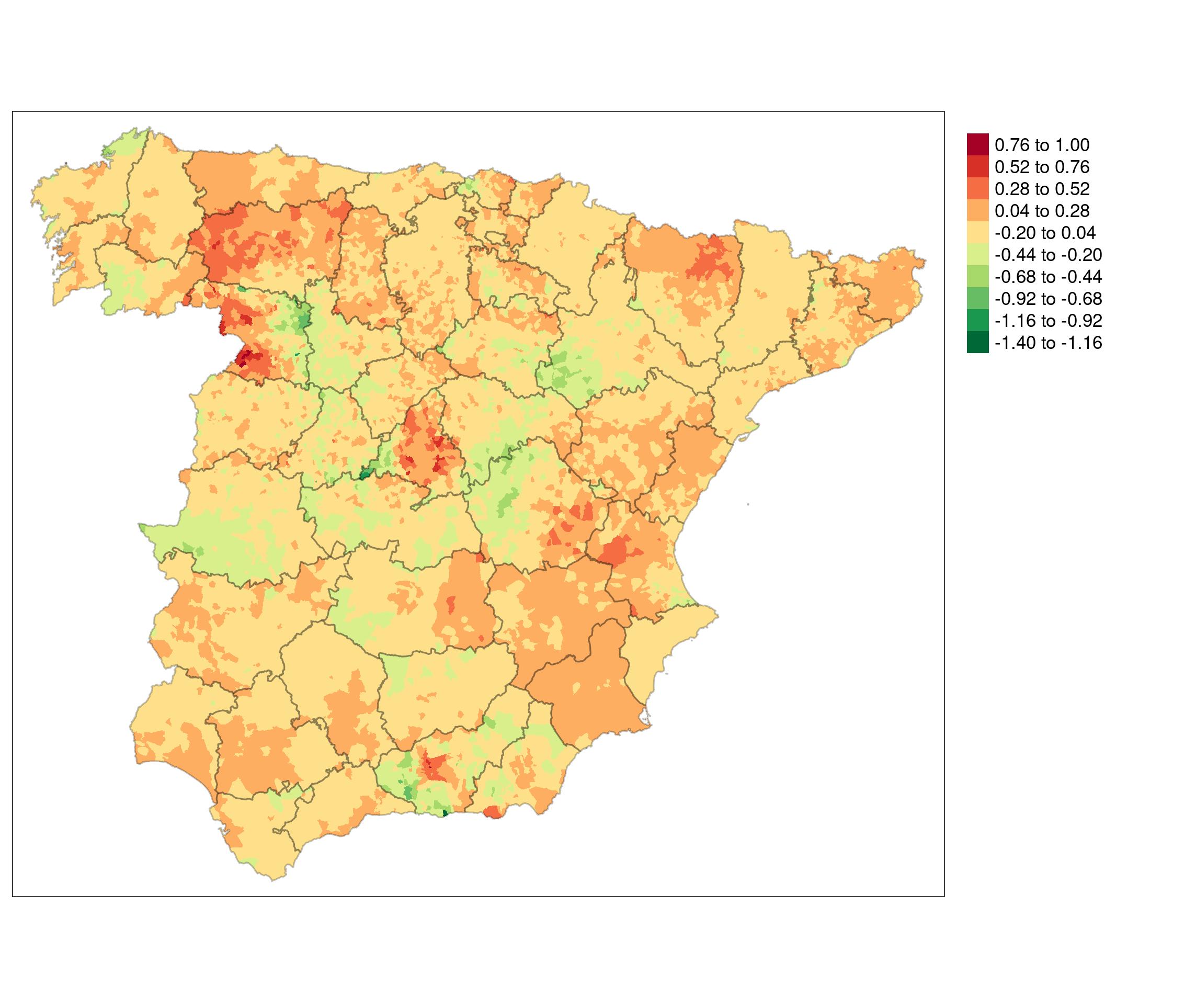}\\
    \vspace{-1cm}
    \includegraphics[scale=0.65]{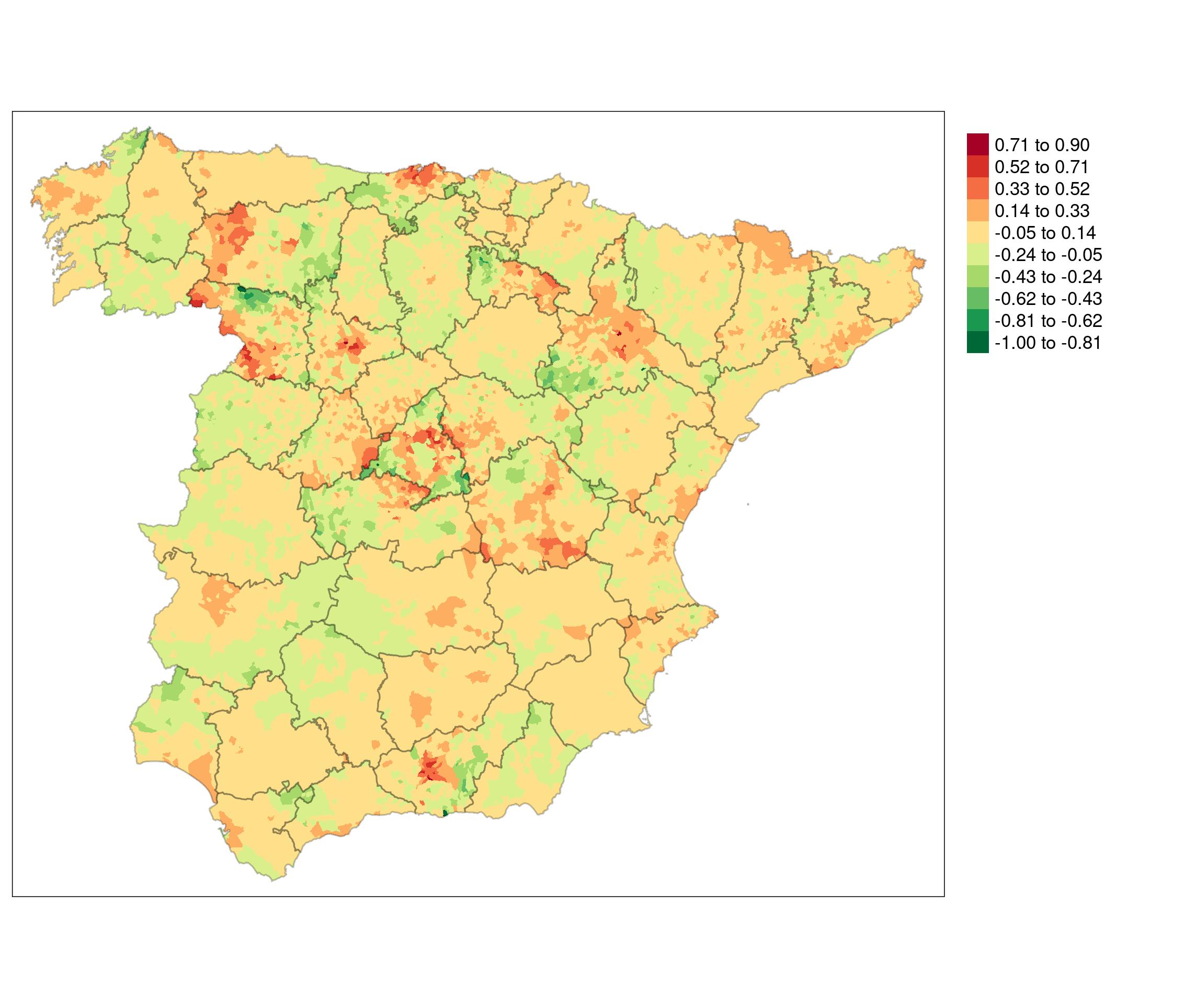}
    \vspace{-1cm}
    \caption{Posterior median of the percentage change from 2013 to 2015 for lung cancer (top) and overall cancer (bottom), i.e.~$100\% \times [(\lambda_{i\text{2015}} -\lambda_{i\text{2013}})/\lambda_{i\text{2013}}]$.}
    \label{fig:Percentage_change}
\end{figure}

In \autoref{fig:Trends_Muni} (bottom) we show the temporal evolution of estimated overall cancer mortality rates for the municipalities of Girona, Madrid and Bilbao. Wider credibility intervals are observed in Bilbao and Girona compared to Madrid, as expected due to the number of inhabitants. Crude rates are always included in the credible intervals.
Finally, \autoref{fig:Percentage_change} (bottom) shows the posterior median of the percentage change in overall cancer mortality rates between 2013 and 2015 at municipality level. It can be seen that the most frequent variations are between $-0.1\%$ and $0.1\%$ in the period 2013-2015. 

\clearpage

\section{Discussion}
\label{sec:Discussion}

Short-term disease forecasting is of great interest in epidemiology and public health as it supports health decision-making processes. 
However, this might be a very complex task when predicting counts for high-dimensional areal data. As far as we know, our paper is the first attempt to extend classical spatio-temporal disease mapping models for predicting short-term cancer burden when the number of  areas is very large. Under this high resolution spatial setting, the main limitation of classical models is its computational complexity due to the huge dimension of the spatio-temporal covariance matrices and the high number of constraints to make the models identifiable.
The divide-and-conquer approach for high-dimensional count data proposed by \cite{orozco2022} is a strategy that involves partitioning the spatial domain into smaller regions and then fitting a separate model for each partition. This approach has been shown to be effective for estimating disease risks, as it can provide better accuracy and computational efficiency compared to classical spatio-temporal disease mapping models. In this paper, we extend this approach to a forecasting setting, where the goal is to forecast short-term mortality rates.

To evaluate the predictive performance of both classical and scalable modelling approaches we perform a validation study imitating the real scenario of forecasting short-term cancer mortality rates in almost $\numprint{8000}$ municipalities of Spain. 
For the partition models, we use the 47 provinces of Spain (NUTS3 level from the European nomenclature of territorial units for statistics) 
although other partitions can be used depending on the researcher's interest. Then, we compute mean absolute errors, root mean square errors and interval scores when making one, two and three-years ahead predictions. In general, partition models outperform the classical models in terms of predicted counts, being the Type IV interaction model the one showing better results. When computing these measures stratifying the data based on the proportion of areas with zero observed cases during the study period or selecting only municipalities lying at the boundary of two or more provinces, we observe that the 1st-order neighbourhood model outperforms the disjoint model in terms of prediction accuracy and interval score. Of note, including 2nd-order neighbourhood models in our validation study do not improve the results further (results are available upon request).
We also illustrate our proposed methodology by forecasting three years ahead lung cancer and overall cancer mortality data in the municipalities of continental Spain using data from the period 1991-2012. To compare the different models in terms of their predictive performance, we compute the logarithmic score based on both leave-one-out (LOOCV) and leave-group-out (LGOCV) cross-validation techniques. Based on these measures, a 1st-order neighbourhood model with Type IV interaction model is chosen as the best model.  
The differences between the models are more pronounced when analysing the data for overall cancer mortality, because the number of
observed cases are higher. Well-known model selection criteria such as the deviance information criterion (DIC) and Watanabe-Akaike information criterion (WAIC) provide similar conclusions.

In summary, the results of this paper show that the divide-and-conquer approach performs well in terms of accuracy and computational time, and outperforms classical methods in all the scenarios. These findings suggest that this approach is a promising strategy for forecasting high-dimensional count data, and can provide valuable insights for decision-making in various fields, such as public health and environmental monitoring.
We remark that the scalable methodology presented in this paper could also be used to forecast rates using other hierarchical Bayesian disease mapping models including, for example, AR terms or second order random walks for time. Moreover, the fact that the proposed methodology is general and can be applied to other health indicators such as cancer incidence or other health indicators is also valuable. This suggests that the methodology can be used for a broader range of applications beyond the scope of the paper.


\section*{Acknowledgements}
This research has been supported by the project PID2020-113125RB- I00/MCIN/AEI/10.13039/501100011033. 

\bibliographystyle{apalike}
\bibliography{biblio}   

\setcounter{section}{0} 
\renewcommand{\thesection}{\Alph{section}}

\setcounter{figure}{0}
\renewcommand\thefigure{\thesection\arabic{figure}}

\setcounter{table}{0}
\renewcommand\thetable{\thesection\arabic{table}}

\clearpage

\section{Appendix: Predictive distribution of the counts and posterior median of percentage change}
\label{sec:AppendixA}

\texttt{R-INLA} provides posterior marginal estimates for the mortality rates $\lambda_{it}$, for $i=1, \ldots, N$ and $t=1, \ldots T$, after setting the arguments \texttt{control.predictor=list(compute=TRUE, link=1)} and \texttt{control.inla=list(return.marginals.predictor=TRUE)} of the main \texttt{inla()}-function call. It also provides predicted distribution estimates of mortality rates $\lambda_{it^{\ast}}$ by setting the observed counts at time point $t^{\ast}$ as \texttt{NA} and giving the corresponding offset $n_{it^{\ast}}$ (in our case, the number of population at risk at municipality $i$ and predicted year $t^{\ast}$).

Using the law of iterated expectation, the expected value for the posterior predictive counts is given by $\mu_{it^{\ast}} = E[E[y_{it^{\ast}} | \lambda_{it^{\ast}}]]= E[n_{it^{\ast}} \cdot \lambda_{it^{\ast}}] = n_{it^{\ast}} \cdot E(\lambda_{it^{\ast}})$. We can also compute the posterior quantiles for the predicted counts by sampling from the marginal posterior of $\lambda_{it^{\ast}}$ \citep{martino2020}.
Our sampling scheme proceeds in two steps. First, we generate $S=\numprint{5000}$ samples from the posterior marginal distribution of $\lambda_{it^{\star}}$ using the function \texttt{inla.rmarginal()} function. Then, we generate values of the mortality counts $y_{it^{\star}}^{s}$ from a Poisson distribution with mean $n_{it^{\star}} \cdot \lambda_{it^{\star}}^{s}$, for $s=1,\ldots,S$, in order to compute its empirical quantiles.

In Section~\ref{sec:DataAnalysis} we present the posterior median of the percentage change of mortality rates from 2013 to 2015. This measure is computed by sampling from the joint posterior distribution of $\lambda_{it}$ using the $\texttt{inla.posterior.sample()}$ function of \texttt{R-INLA}, and then, computing for each municipality the median value of the ratios $100 \times [(\lambda_{i\text{2015}}^{s} -\lambda_{i\text{2013}}^{s})/\lambda_{i\text{2013}}^{s}]$ for $s=1,\ldots,\numprint{10000}$.

\bigskip
\section{Appendix: additional figures}
\label{sec:AppendixB}

\begin{figure}[!ht]
    \centering
    \includegraphics[scale=0.35]{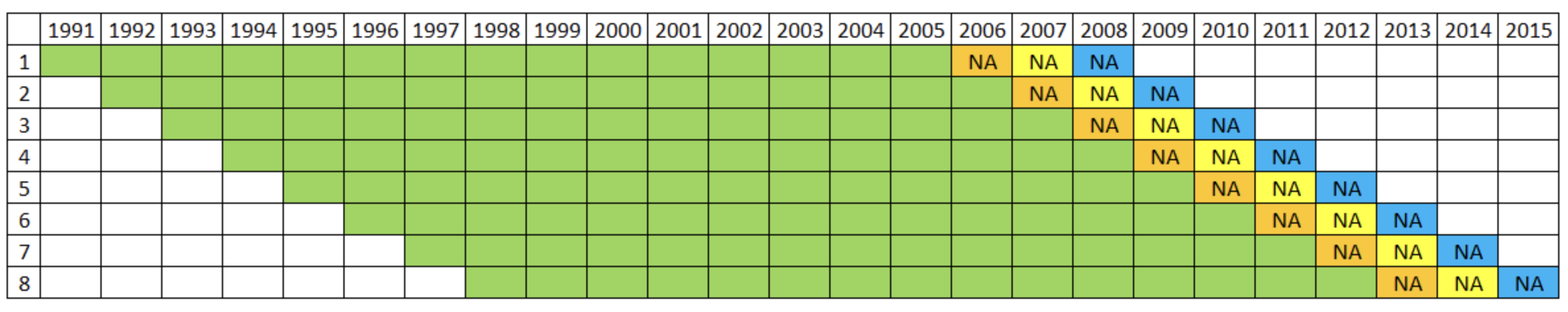}
    \caption{Validation setup: Each row corresponds to one of the eight validation configurations. Green cells represent years with data that are used in the model. Orange, yellow, and blue cells indicate years for which one, two, and three years ahead predictions are computed, respectively.}
    \label{fig:cascada}
\end{figure}

\clearpage
\section{Appendix: additional tables}
\label{sec:AppendixC}

\begin{table}[!ht]
\caption{Average values of prediction evaluation scores for models fitted with INLA (simplified Laplace approximation strategy) in the municipalities with proportion of zero observed cases during the study period less or equal than 0.2.}
\label{tab:ValidationStudy_PropZero}
\renewcommand{\arraystretch}{1.1}
\begin{center}
\resizebox{\textwidth}{!}{
\begin{tabular}{llrrrrrrrrrrrrr}
\hline
\multirow{2}{20mm}{Model} & \multirow{2}{20mm}{Space-time interaction} & \multicolumn{3}{c}{1-year ahead} & & \multicolumn{3}{c}{2-years ahead} & &\multicolumn{3}{c}{3-years ahead} \\
\cline{3-5} \cline{7-9} \cline{11-13}
& & IS$_{0.05}$ & MAE & RMSE & & IS$_{0.05}$ & MAE & RMSE & & IS$_{0.05}$ & MAE & RMSE & \\
Classical     & Type I   & 12.96 & 2.20 & 2.67 & & 13.22 & 2.25 & 2.72 & & 13.18 & 2.30 & 2.76 \\
              & Type II  &   $-$ &  $-$ &  $-$ & &   $-$ &  $-$ &  $-$ & &   $-$ &  $-$ &  $-$ \\
              & Type III & 13.10 & 2.21 & 2.68 & & 13.14 & 2.26 & 2.73 & & 13.50 & 2.31 & 2.78 \\
	          & Type IV  &   $-$ &  $-$ &  $-$ & &   $-$ &  $-$ &  $-$ & &   $-$ &  $-$ &  $-$ \\
\hline
Disjoint      & Type I   & 12.93 & 2.18 & 2.65 & & 13.09 & 2.23 & 2.70 & & 13.21 & 2.27 & 2.74 \\
by provinces  & Type II  & 12.65 & 2.13 & 2.59 & & 12.93 & 2.16 & 2.63 & & 13.25 & 2.21 & 2.67 \\
              & Type III & 12.76 & 2.17 & 2.64 & & 12.92 & 2.22 & 2.69 & & 13.08 & 2.26 & 2.72 \\
              & Type IV  & \textbf{12.51} & \textbf{2.12} & \textbf{2.59} & & \textbf{12.79} & \textbf{2.17} & \textbf{2.64} & & \textbf{13.09} & \textbf{2.21} & \textbf{2.68} \\
\hline
1st order     & Type I   & 12.88 & 2.17 & 2.64 & & 13.02 & 2.21 & 2.69 & & 13.14 & 2.26 & 2.73 \\
neighbourhood & Type II  & 12.57 & 2.13 & 2.58 & & 12.88 & 2.16 & 2.63 & & 13.18 & 2.21 & 2.67 \\
by provinces  & Type III & 12.73 & 2.16 & 2.63 & & 12.92 & 2.21 & 2.68 & & 12.96 & 2.25 & 2.71 \\
              & Type IV  & \textbf{12.48} & \textbf{2.12} & \textbf{2.58} & & \textbf{12.72} & \textbf{2.16} & \textbf{2.63} & & \textbf{12.96} & \textbf{2.21} & \textbf{2.68} \\
\hline
\end{tabular}}
\end{center}
\end{table}

\begin{table}[!ht]
\caption{Average values of prediction evaluation scores for models fitted with INLA (simplified Laplace approximation strategy) in the municipalities lying at the boundary between two or more provinces with at least 2 observed cases per $\numprint{100000}$ inhabitants during the whole study period.}
\label{tab:ValidationStudy_Boundary}
\renewcommand{\arraystretch}{1.1}
\begin{center}
\resizebox{\textwidth}{!}{
\begin{tabular}{llrrrrrrrrrrrrr}
\hline
\multirow{2}{20mm}{Model} & \multirow{2}{20mm}{Space-time interaction} & \multicolumn{3}{c}{1-year ahead} & & \multicolumn{3}{c}{2-years ahead} & &\multicolumn{3}{c}{3-years ahead} \\
\cline{3-5} \cline{7-9} \cline{11-13}
& & IS$_{0.05}$ & MAE & RMSE & & IS$_{0.05}$ & MAE & RMSE & & IS$_{0.05}$ & MAE & RMSE & \\
\hline
Classical     & Type I   & 16.14 & 2.62 & 3.20 & & 16.40 & 2.66 & 3.25 & & 15.97 & 2.65 & 3.23 \\
              & Type II  &  $-$ &    $-$ &    $-$ & &  $-$ &    $-$ &    $-$ & &  $-$ &    $-$ & $-$ \\
              & Type III & 16.36 & 2.62 & 3.20 & & 16.01 & 2.67 & 3.26 & & 16.01 & 2.66 & 3.23 \\
              & Type IV  &  $-$ &    $-$ &    $-$ & &  $-$ &    $-$ &    $-$ & &  $-$ &    $-$ & $-$ \\
\hline
Disjoint      & Type I   & 15.82 & 2.61 & 3.15 & & 15.92 & 2.65 & 3.20 & & 15.79 & 2.64 & 3.18 \\
by provinces  & Type II  & 15.87 & 2.63 & 3.18 & & 16.07 & 2.64 & 3.21 & & 16.19 & 2.64 & 3.19 \\
              & Type III & 15.58 & 2.61 & 3.15 & & 15.79 & 2.65 & 3.20 & & 15.77 & 2.64 & 3.19 \\
              & Type IV  & \textbf{15.71} & \textbf{2.62} & \textbf{3.17} & & \textbf{15.98} & \textbf{2.65} & \textbf{3.20} & & \textbf{15.99} & \textbf{2.64} & \textbf{3.19} \\
\hline
1st order     & Type I   & 15.70 & 2.60 & 3.15 & & 15.67 & 2.64 & 3.21 & & 15.54 & 2.64 & 3.19 \\
neighbourhood & Type II  & 15.70 & 2.62 & 3.17 & & 15.69 & 2.64 & 3.21 & & 15.83 & 2.63 & 3.18 \\
by provinces  & Type III & 15.50 & 2.60 & 3.15 & & 15.50 & 2.65 & 3.22 & & 15.36 & 2.65 & 3.20 \\
              & Type IV  & \textbf{15.60} & \textbf{2.60} & \textbf{3.15} & & \textbf{15.63} & \textbf{2.64} & \textbf{3.20} & & \textbf{15.62} & \textbf{2.63} & \textbf{3.18} \\
\hline	
\end{tabular}}
\end{center}
\end{table}

\end{document}